\documentclass[10pt,preprint]{article}
\usepackage{osajnl2}

\begin{document}

\title{Analytical theory of dark nonlocal solitons}

\author{Qian Kong$^{1,2}$, Q. Wang$^{2}$, O. Bang$^3$, W. Krolikowski$^1$}

\address{$^1$Laser Physics Center, Research School of Physics and Engineering,
Australian National University, Canberra ACT 0200, Australia\\
$^2$Department of Physics, Shanghai University, Shanghai 200444, China\\
$^3$DTU Fotonik, Technical University of Denmark, Lyngby, Denmark}

\email{} 



\begin{abstract}
We investigate properties of dark solitons in nonlocal materials with
an arbitrary degree of nonlocality. We employ the variational technique
and describe the dark solitons, for the first time, in the whole range of degree of nonlocality.
 \end{abstract}

 

Spatial dark optical solitons are localized light intensity dips
in the infinite  constant
background~\cite{Swartzlander:prl:91,Kivshar:pr:98,Burger:prl:99}.
Like their bright counterparts, they propagate preserving their
spatial profile. This is a result of a balance between diffraction
 which tends to broaden their size and nonlinearity of the medium which counteracts this process.
The existence  of dark spatial solitons  requires the nonlinearity of the medium
 to be negative, or self-defocusing, i.e. refractive index of the medium must decrease with
 light intensity~\cite{Swartzlander:prl:91}. In the simplest case  of Kerr medium  the change of the
 refractive  index is just proportional to the light intensity. In this special situation and in one-dimensional case
 the soliton amplitude $u(x,z)$  is described analytically by the following relation
 \begin{equation}
 u(x,z)=B\tanh(x-vz)+iA,
 \end{equation}
 where $x$ and $z$ denote the transverse and propagation coordinates, respectively. B and A are constant and $v$ is soliton transverse velocity, $v=A/B$. In case of more complex relation between light intensity and the nonlinearity the
 solitons profiles, except for some  special models,  can be only found numerically~\cite{Kivshar:pr:98}.
 All the typical
 nonlinear models are   {\em local} i.e. the nonlinear response  in a particular point  is determined solely by the light intensity in the same point.  However, there has been  recently strong interest in the so called nonlocal nonlinearities~\cite{wk:job:04}. In those models the relation between nonlinear response and the intensity is spatially nonlocal. Physically it  means that the nonlinearity in a particular spatial location is determined by the light intensity in a certain neighborhood of this location. Typical nonlocal systems involve, either transport processes such as heat~\cite{Dabby68} or ballistic atomic transport~\cite{Skupin07:prl} and  diffusion~\cite{Suter93}, charge separation~\cite{Ultanir03ol} or long-range interaction as in dipolar Bose Einstein condensate~\cite{Dalfovo99} or nematic liquid crystals~\cite{Braun93}.
 Nonlocal nonlinearity has been shown to have a stabilizing effect on nonlinear structures ~\cite{Bang02,wk:job:04,lin:josab:08,andrea:pra:09} enabling formation of complex solitonic entities such as multi-pole solitons or azimuthons~\cite{Skupin:pre:06}. It also affects soliton interaction inducing a long-range attractive forces between solitons~\cite{SnyderMitchell:science:97,Peccianti:ol:02,Rasmussen:pre:05}.
 In case of self-defocusing nonlinearities, nonlocality has been shown to modify soliton interaction leading to attraction of dark solitons and formation of their bound states~\cite{Nikolov:ol:04,Dreischuh:prl:06}.
 Unlike the Kerr nonlinear models where the solitons can be described analytically, fully nonlocal models with arbitrary degree of nonlocality can only be  treated numerically~\cite{Kartashov:ol:07,Guo:oex:09}.  The only exceptions are   the special case of the so called weak and strong nonlocality, when  the steady state bright and dark soliton could be found in analytical form~\cite{wkob:pre:01,Nikolov:pre:03}.

 In this work we describe analytically, for the first time,   properties of dark nonlocal solitons.
 To this end we employ a  variational approach and show  that it enables one
 to retrieve the major features  of dark  solitons in a general nonlocal regime.


The evolution of one-dimensional optical beam with an amplitude $u(x,z)$ in
nonlocal defocusing medium is governed by the following  nonlocal nonlinear Schrodinger equation (NLS) ~\cite{Kivshar:pr:98}

\begin{equation}\label{NLS}
i\frac{\partial u}{\partial z} + \frac{1}{2}\frac{\partial^{2}
u}{\partial x^{2}} - u\int^{+\infty}_{-\infty}
R(x-\xi)|u(\xi,z)|^2 d\xi = 0.
\end{equation}
 Here we used the phenomenological model of the nonlocal nonlinearity   with  $R(x)$ being  the nonlocal response function. Its width determines the degree of nonlocality~\cite{wkob:pre:01}. In particular, for  $R(x)=\delta(x)$ the above equation describes standard Kerr local medium. While this is only a phenomenological model it nevertheless describes very well general properties of the nonlocal media. Moreover, for certain form of the nonlocal response function this model represents the actual physical system. For instance, this is exactly the case of nematic liquid crystals, when the long-range interaction-mediated nonlinear response function, under certain approximations~\cite{Conti:prl:03}, can be written as $R(x)\propto \exp {\left(-|x|/\sigma\right )}$ with $\sigma$ being the degree of nonlocality. Also, the same exponential response describes nonlinear interaction in quadratic media~\cite{Nikolov:pre:03}.

To analyze the nonlocal NLS equation,  we will employ the Lagrangian approach which in case of dark solitons has been formulated in \cite{Kivshar:oc:95}.  Following this work we find that the renormalized Lagrangian density corresponding to the NLS Eq.(\ref{NLS}) is given in the following form
\begin{equation}\label{density}
\mathcal{L}=\frac{i}{2}\left(u^{*}\frac{\partial u}{\partial
z}-u\frac{\partial u^{*}}{\partial
z}\right)\left(1-\frac{1}{|u|^{2}}\right)-\frac{1}{2}\left|\frac{\partial
u}{\partial
x}\right|^{2}-\frac{1}{2}(|u|^{2}-1)\int^{\infty}_{-\infty}
R(x-\xi)(|u(\xi,z)|^{2}-1)d\xi.
\end{equation}

To proceed further we need to  specify  the nonlocality.
 For the sake of simplicity and analytical tractability and without the loss of generality
  we consider here   the  rectangular profile for the nonlocal response function~\cite{wkob:pre:01}  $R(x)$:
\begin{equation}\label{response}
R(x)=\left\{\begin{array}{ll} \frac{1}{2\sigma} & -\sigma \leq x \leq \sigma,\\
0 & \mbox{otherwise.}
\end{array}\right.,
\end{equation}
with  $\sigma$  defining  the  degree of nonlocality. For
$\sigma\rightarrow 0$ we get $R(x)\rightarrow \delta(x)$ and the
model describes just  the local Kerr nonlinearity. The accuracy of
the Lagrangian approach depends on the functional choice of the
variational solution. Here  we use the obvious ansatz
\begin{eqnarray}\label{ansatz}
u(x,z) &=& B \tanh[D(x-x_{0})]+i A,\nonumber \\ A^{2}+B^{2} &=& 1,
\end{eqnarray}
which  represents an  exact soliton solution in the local regime.
All  parameters A,B,D, and $x_{0}$ are assumed to be
functions of the propagation variable $z$. Substituting Eq.(\ref{ansatz}) and
Eq.(\ref{response}) into the  Lagrangian density Eq.(\ref{density})
and integrating over $x$, we get the "averaged"
Lagrangian,

\begin{equation}
L = \int^{\infty}_{-\infty}\mathcal{L}(u)dx  = 2\frac{d x_{0}}{dz}\left[-AB
+\tan^{-1}\left(\frac{B}{A}\right)\right]-\frac{2}{3}B^{2}D+\frac{B^{4}}{D}\left[\mbox{csch}^{2}(D\sigma)
-\frac{\coth(D\sigma)}{D\sigma}\right].
\end{equation}

Then we find the  corresponding Euler-Lagrange equations  as

\begin{eqnarray}\label{nonlocal_solution}
\frac{dB}{dz} = 0 , \nonumber \\
\frac{\coth(D\sigma)}{D\sigma}\left[\frac{1}{D^{2}}-\sigma^{2}\mbox{csch}^{2}(D\sigma)\right] = \frac{1}{3B^{2}},  \nonumber \\
\frac{d x_{0}}{dz} =
A\left[\frac{D}{3B}-\frac{B}{D}\left(\mbox{csch}^{2}(D\sigma)-\frac{\coth(D\sigma)}{D\sigma}\right)\right].
\end{eqnarray}

Notice that for $\sigma=0$ the above formulas give $B=D=const$ and $dx_0/dz=A$ which are the exact soliton parameters for local Kerr solitons~\cite{Kivshar:oc:95}.
The formulas Eq.(\ref{nonlocal_solution}) constitute the main result of this paper. They give, for the first time, the analytical relation among parameters of the dark nonlocal soliton, for arbitrary degree of nonlocality. In particular, one can use it to show the dependence of the soliton width ($\propto D^{-1}$) as a function of the degree of nonlocality ($\sigma$). This relation is depicted in Fig.1 by the solid line. Here the soliton width is normalized to its value in the local regime. The graph clearly shows that the width of dark soliton is a nonmonotonic function of the nonlocality. It decreases first for small $\sigma$,  reaches its minimum and then monotonically increases. This nontrivial analytical result agrees  with that found earlier in numerical simulations [see Fig.3 in ~\cite{Nikolov:pre:03}].
On the physical grounds the initial decrease of the soliton width can be explained
by the fact that  weak nonlocality causes the nonlinear index change to advance towards regions of lower light intensity. As a result the waveguide induced by the soliton becomes slightly narrower and so does the soliton. Situation becomes different for high degree of nonlocality. For large $\sigma$ the refractive index modulation expressed by the convolution integral in Eq.(\ref{NLS}) becomes weaker   and broader, acquiring  wide rectangular  profile and resulting in increased width of the soliton.

\begin{figure}
\includegraphics[width=1.0\columnwidth]{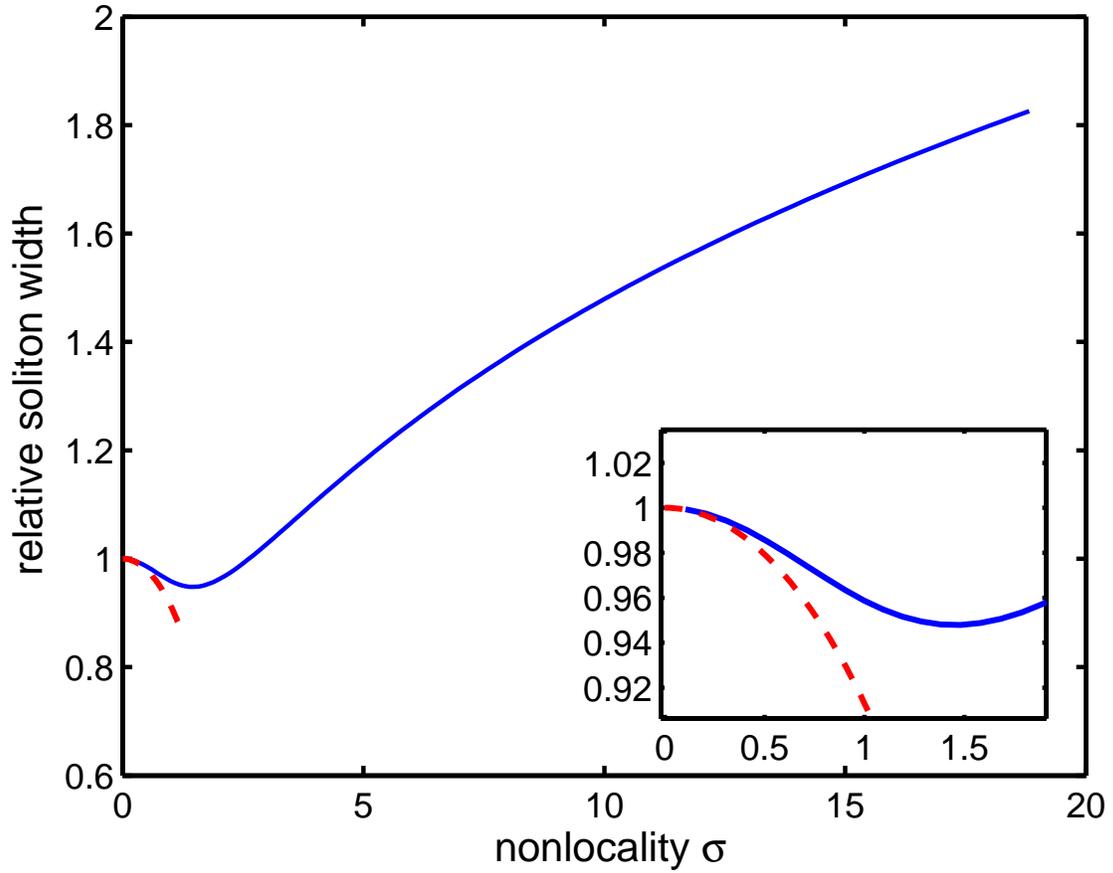}
\caption{\label{fig1}Width of the dark nonlocal soliton (normalized to its value in the local limit)
as a function of the degree of nonlocality $\sigma$.
solid (blue) line - variational results with rectangular response function; dashed (red) line - exact solution in a weakly nonlocal limit Ref.~\cite{wkob:pre:01}. The inset depicts the magnified region of weak nonlocality. Here $B\approx 1$, $A\approx 0$}
\end{figure}

It is instructive to compare the above  variational calculations with the exact analytical
solutions which can be obtained in two limiting regimes  of weak and strong nonlocality. In the former case
when the response function is much narrower than the soliton width, the convolution term in
Eq.(\ref{NLS}) can be expanded in Taylor series leading to the following form of the nonlinear response
\begin{equation}
\int_{-\infty}^{+\infty} R(x-\xi)|u(\xi,z)|^2 d\xi \approx
|u(x)|^2+\gamma \frac{\partial^2|u(x)|^2}{\partial x^2},
\end{equation}
where we introduced parameter  $\gamma=\sigma^2/6$.
The NLS with such nonlinear term can be solved analytically as shown in Ref.~\cite{wkob:pre:01}.
The weakly nonlocal limit can be  recovered from
the general variational solutions Eq.(\ref{nonlocal_solution})
  by  expanding  it into a Taylor series around $\sigma=0$. After retaining
  only the most significant terms   we obtain
\begin{eqnarray}
\frac{dB}{dz} &=& 0, \nonumber \\
D^{2} &=& \frac{B^{2}}{1-\frac{2}{15}\sigma^{2}B^{2}} \nonumber \\
\frac{d x_{0}}{dz} &=&
A\left[\frac{1}{3B}\left(D+\frac{2B^{2}}{D}\right)-\frac{4\sigma^{2}BD}{45}\right],
\end{eqnarray}
which coincides exactly with the  equations derived
directly  from the Lagrangian representation of the weakly nonlocal nonlinear Schr\"{o}dinger equation~\cite{Wang:jopt:09}
\begin{figure}
\includegraphics[width=1.0\columnwidth]{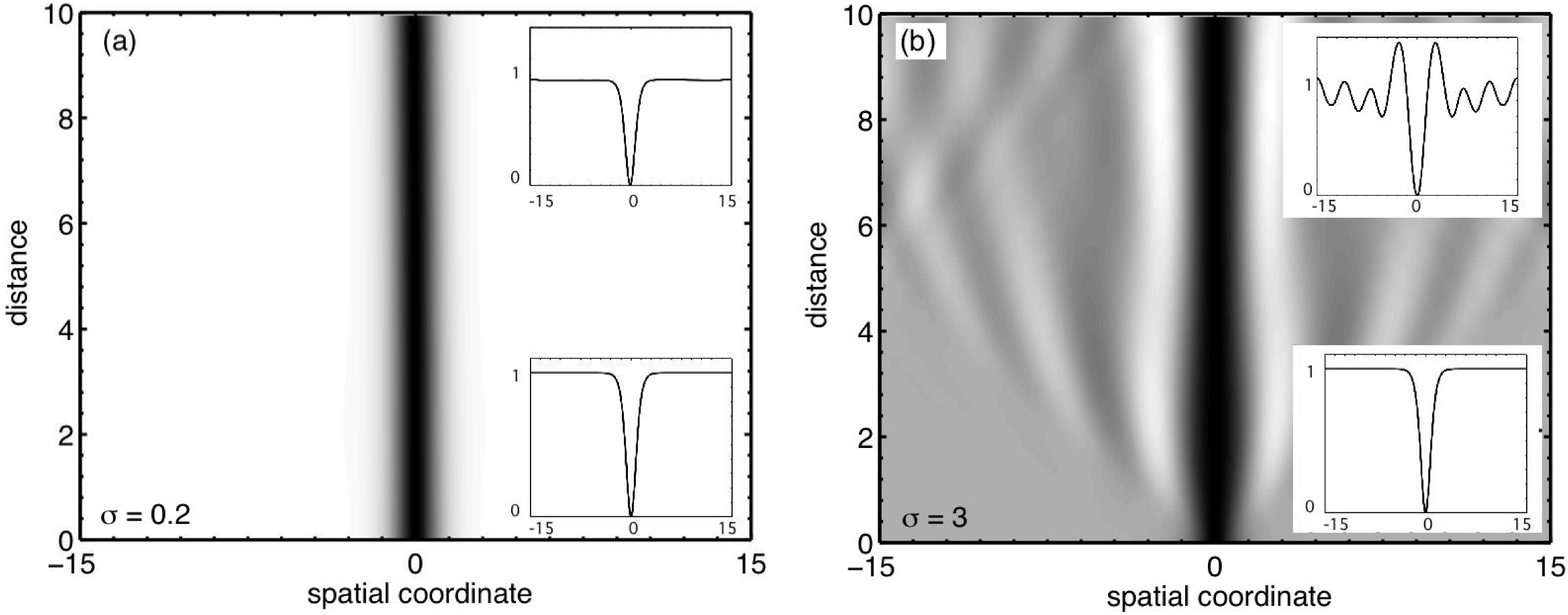}
\caption{\label{fig2} Dynamics of formation of dark spatial solitons in (a) almost local, $\sigma=0.2$ and (b) nonlocal, $\sigma=3$,  media with rectangular response function. The insets depict the soliton profile at the beginning (bottom) and end (top) of the propagation distance.}
\end{figure}
The exactly found width of dark soliton in a  weakly nonlocal regime solution is shown in Fig.1  by  dashed (red) line. It is evident that the agreement between variational and exact solutions is indeed very good. This is an obvious consequence of the fact that the ansatz, Eq.(\ref{ansatz}), is actually pretty close to the exact solution in this regime.
 Different situation arises in the so called highly nonlocal limit. It has been shown that in such limit the nonlocal  NLS becomes linear with the convolution term being just proportional to the response function.
 In this regime this equation can be solved exactly. For the rectangular response function the solution is
 \begin{equation}\label{highly_nonlocal}
 u(x)=B\sin\left (\frac{\pi(x-x_0)}{2\sigma}\right )+iA
 \end{equation}
 This gives monotonic increase of the soliton width with nonlocality $D^{-1}\propto \sigma$. This of course differs from the variational solution Eq.(\ref{nonlocal_solution}) which for large degree of nonlocality gives $D^{-1}\propto \sigma^{1/3}$. This discrepancy is most likely caused by the inadequacy of our ansatz in the highly nonlocal regime as will be evident below.

In Fig.2 we illustrate the effect of nonlocality on the formation
and propagation of single dark spatial soliton. We used split step
Fourier method to integrate numerically the nonlinear Schrodinger
equation. Fig.2 depict dynamics of dark soliton excited by the
initial conditions represented by the variationally calculated
profiles with $\sigma=3$. Graphs  (a) and (b) correspond to almost
local ($\sigma=0.2$) and nonlocal ($\sigma=3$) cases,
respectively. It is clear that nonlocality tends to expand the
width of soliton. The strong dispersive waves visible in graph (b)
reflect the fact that the initial conditions do not correspond to
the stationary soliton profile. This is obvious after noticing
that unlike the local case when the soliton intensity
monotonically grows from its minimum   to its background value,
the nonlocal intensity profile almost always exhibit strong
overshot [clearly visible in the inset in Fig.2(b)].
Unfortunately, the ansatz Eq.(\ref{ansatz}) does not reflect this
property and hence excitation of dark soliton using our
variational  solutions is always accompanied by the emission of
diffractive waves.

In conclusion, we studied properties of  single dark solitons in nonlocal nonlinear media. We used the variational approach an the rectangular model of the nonlocality to derive, for the first time,  analytical relations for  soliton parameters for arbitrary degree of nonlocality. We showed that this approach, while approximate, faithfully represent properties  of dark nonlocal solitons.

This work was supported by the National Natural Science Foundation
of China (Grant No. 60808002), the Shanghai Leading Academic
Discipline Program (Grant No. S30105), the China Scholarship
Council and the Australian Research Council.


\end{document}